\documentclass[aps,prl,reprint,superscriptaddress]{revtex4-2}
\usepackage{slashed}
\usepackage{amsmath}
\usepackage{amssymb}
\usepackage{graphicx}
\usepackage{mathtools}
\usepackage[caption=false,position=top]{subfig}
\usepackage[export]{adjustbox}
\usepackage{dsfont}
\usepackage{xcolor}
\usepackage{comment}
\usepackage{lineno}
\usepackage{hyperref}
\hypersetup{colorlinks=true,linkcolor=black,citecolor=blue,urlcolor=blue}

\def\dd{\mathrm{d}}
\def\be{\begin{equation}}
\def\ee{\end{equation}}
\def\bea{\begin{eqnarray}}
\def\eea{\end{eqnarray}}

\begin{document}

\title{`Stealth' singularities from self-gravitating fermions}

\author{Peter E.~D.~Leith}
\affiliation{SUPA, School of Physics and Astronomy, University of St Andrews, North Haugh, St Andrews, Fife KY16 9SS, UK}
\author{Chris A.~Hooley}
\affiliation{Centre for Fluid and Complex Systems, Coventry University, Coventry CV1 2TT, UK}
\author{Keith Horne}
\affiliation{SUPA, School of Physics and Astronomy, University of St Andrews, North Haugh, St Andrews, Fife KY16 9SS, UK}
\author{David G.~Dritschel}
\affiliation{School of Mathematics and Statistics, University of St Andrews, North Haugh, St Andrews, Fife KY16 9SS, UK}

\date{\today}

\begin{abstract}
We present a new analytic solution to the Einstein--Dirac equations formulated by Finster, Smoller, and Yau [{\it Phys.\ Rev.\ D\/}\ {\bf 59}, 104020 (1999)] to describe the stationary states of a pair of gravitationally interacting neutral fermions.  The fermions' wavefunction in our analytic solution, as in their numerical ones, is both exponentially localized and normalizable.  However, our solution differs from theirs in two key respects:\ it features a naked spacetime singularity at the origin, and the gravitational (Arnowitt--Deser--Misner) mass of the localized object is zero, making it gravitationally undetectable to an external observer. This is despite the arbitrarily large mass of the constituent fermions. This unexpected result may have significant implications for astronomy and cosmology, as it gives a mechanism by which mass could become ‘hidden’ during the universe’s evolution.
\end{abstract}

\maketitle

{\it Introduction.} The existence of so-called `naked' singularities (those without an accompanying horizon) in general relativity has been debated for many years~\cite{JoshiSing}. It is clear that the Einstein equations admit solutions of this type, but the consensus is that naked singularities do not form as a result of gravitational collapse, at least not in an astrophysical context. This is now supported by the recent observational evidence of black holes~\cite{EHTcollab} and black-hole mergers~\cite{Abbott2016}. Furthermore, it is generally accepted that the matter sector associated with spacetimes containing naked singularities must exhibit a degree of unphysicality, such as in the negative-mass Schwarzschild solution~\cite{Bondi1957}, thus limiting their real-world relevance. 

In the quantum regime, however, the matter sector is not necessarily subject to the same stringent physicality constraints~\cite{Belletete2013}, and therefore the existence of naked singularities at the quantum scale is still very much an open question. It is possible that such objects could have relevance in the context of the early universe, for instance as alternatives to primordial black holes~\cite{joshi2024}. 

In this Letter, we demonstrate that spacetimes containing naked singularities can occur in a simple (semi-classical) quantum system:\ a spinor field minimally coupled to gravity via the Einstein equations. The resulting spacetimes are asymptotically flat, but intriguingly we find that the total gravitational mass, measured e.g.\ by the Arnowitt--Deser--Misner (ADM) mass, is identically zero, regardless of the mass of the spinor field, implying that these objects would be gravitationally undetectable by a test particle situated at a sufficient distance (beyond the order of a few Planck lengths). This raises the possibility of effectively concealing a clump of fermionic matter from external observation, assuming that the only interaction is via gravity. 

Soliton-like objects with this property have previously been found in the context of gravitationally interacting skyrmions~\cite{Klinkhamer2018stealth,Klinkhamer2019lensing,Klinkhamer2019soliton}, and termed ``stealth defects''. Given the similarities between the two cases, we describe the objects presented here as ``stealth singularities''.

Compared to the Skyrme model, the system studied here is relatively simple, being composed of the minimally coupled Dirac and Einstein equations (together referred to as the Einstein--Dirac system):
\begin{align}
	&\left(\slashed{D}-m\right)\Psi_a=0;
	&&G_{\mu\nu}=8\pi G\, T_{\mu\nu}\left[ \left\{ \Psi_a \right\} \right], \label{eqDEandEE}
\end{align}
where $\slashed{D}$ is the Dirac operator in curved spacetime, $m$ is the fermion mass, $\Psi_a$ is the spinor wavefunction for fermion $a$, $G_{\mu\nu}$ is the Einstein tensor, $G$ is the Newton constant, and $T_{\mu\nu}$ is the stress-energy tensor. This is by no means a full theory of quantum gravity, since the gravitational field is not quantized and the matter sector is not treated as a quantum field.  Nevertheless, this formalism provides an interesting testing ground in which to study the interaction between fermionic quantum particles and gravity in a semi-classical context.

An interesting feature of the Einstein--Dirac system is the existence of `particle-like' solutions, where the gravitational self-interaction of the spinor wavefunction acts to create a spatially localized object, which is prevented from collapse by the effects of the uncertainty principle. Generally referred to as either `Dirac solitons' or `Dirac stars', these objects were first studied by Finster {\it et al.}~\cite{FSY1999original}, henceforth FSY, who numerically constructed localized, spherically symmetric solutions containing two neutral fermions. This initial analysis has since been extended to solutions that include additional fields~\cite{FSY1999maxwell,FSY2000nonAbelianBound,Leith2023Higgs} and larger numbers of fermions~\cite{Leith2020fermionTrapping,Leith2021excited}, and has also been generalized to axisymmetric one-particle states~\cite{Herdeiro2019bosonDiracProcaSpinning}.

In addition to these numerical studies, a number of analytic solutions to the (spherically symmetric) Einstein--Dirac system have also been discovered. These include a ``power-law solution'' or ``light-like singularity", a Schwarzschild-like black hole solution, and both massive and massless ``wormhole" solutions~\cite{Bakucz2020powerlaw, Blazquez2020ansatz}. Although interesting in their own right, the physical relevance of these solutions is doubtful, since none of them contains a matter sector that is normalizable (due to divergences in the spinor fields either at spatial infinity or at a horizon). 

In contrast, the analytic solution presented here has spinor fields that decay exponentially at large radii, and although the spacetime is singular at $r=0$, the integral over the fermion density nonetheless converges, allowing the spinor fields to be consistently normalized. In this regard, our solution is similar in spirit to the `particle-like' solutions found numerically by FSY:\ it has a normalized, fermionic quantum wavefunction that is spatially localized, with an associated spacetime that is asymptotically flat. There is one key difference, however:\ the presence of a central naked singularity, which significantly alters the properties of the resulting soliton-like objects.

\textit{Equations of motion.} The equations of motion governing the two-fermion Einstein--Dirac system were first derived by FSY; we begin by outlining that derivation.  The spacetime metric $g_{\mu\nu}$ and spinor wavefunctions $\Psi_{a\in\{1,2\}}$ take the following spherically symmetric forms: 
\begin{align}
	&g_{\mu\nu}=\mathrm{diag}\left(-\frac{1}{T(r)^2},\frac{1}{A(r)},r^2,r^2\sin^2\theta\right);\label{metricans}\\
	&\Psi_a(r,t)=\frac{\sqrt{T(r)}}{r}\binom{\alpha(r)e_a}{i\sigma^r\beta(r)e_a}e^{-i\omega t}\,,\label{eq2spinorAns2F}
\end{align}
where $e_1=(1,0)^\mathrm{T}$, $e_2=(0,1)^\mathrm{T}$, $\sigma^r$ is the radial spherical Pauli matrix, $\omega$ is the fermion energy (assumed common for each fermion), and $(t,r,\theta,\phi)$ are the usual spherical co-ordinates. We use natural units $\hbar=c=1$; the Newton constant $G$ is retained explicitly. The ansatz (\ref{eq2spinorAns2F}) represents a pair of fermions of opposite spin arranged in a singlet state.  The two fields in (\ref{metricans}), $A$ and $T$, describe the radial profile of the spacetime metric; the two in (\ref{eq2spinorAns2F}), $\alpha$ and $\beta$, describe the fermion and anti-fermion components of the spinor wavefunction.

Using these ansatzes, one can derive expressions for the Dirac operator $\slashed{D}$, the energy-momentum tensor $T_{\mu\nu}$, the Ricci tensor $R_{\mu\nu}$, and the Ricci scalar $\mathcal{R}$. Substituting these into (\ref{eqDEandEE}), one obtains the following set of four coupled differential equations:
\begin{align}
	\sqrt{A}\,\alpha'&=+\frac{\alpha}{r}-(\omega T+m )\beta\label{eqDE1twoF}\,;\\
	\sqrt{A}\,\beta'&=-\frac{\beta}{r}+(\omega T-m )\alpha\label{eqDE2twoF}\,;\\
	-1+A+rA'&=-16\pi G\omega T^2\left(\alpha^2+\beta^2\right)\label{eqEE1twoF};\\
	-1+A-2rA\frac{T'}{T}&=16\pi G T\sqrt{A}\left(\alpha\beta'-\alpha'\beta\right)\label{eqEE2twoF}.
\end{align}
Here, $'\equiv \dd/\dd r$, and the fermion mass $m$ is common for each fermion. Note that these are valid strictly for fermions with positive parity; negative parity solutions are discussed in the supplemental material, along with the generalization to fermion number $N>2$.

For localized solutions we require the spacetime to be asymptotically flat:
\begin{gather}
	\lim_{r\to\infty}A(r)\to 1; \hspace{20pt}\lim_{r\to\infty}T(r)\to 1.\label{eqAsymFlat}
\end{gather}
We also require each spinor wavefunction to be correctly normalized. This effectively quantizes the system by ensuring that each of the two spinor fields is occupied by precisely one fermion, and implies that
\begin{align}
	\int_0^\infty \frac{4\pi r^2\, \dd r}{\sqrt{A}} n_f = N = 2\label{eqNormOv},
\end{align}
where $r^2n_f = NT(\alpha^2+\beta^2)$ is the (radial) fermion density.



\textit{The stealth solution.} The conditions of asymptotic flatness and normalization, along with (\ref{eqDE1twoF})--(\ref{eqEE2twoF}), fully define the two-fermion Einstein--Dirac system. For the special case of zero fermion energy ($\omega=0$), together with a spatially flat metric ($A=1$), we find that the equations of motion admit an exact solution that is both asymptotically flat and normalizable.  Setting $\omega=0$ and $A=1$, and defining $x \equiv mr$, the equations of motion become:
\begin{align}
	\alpha'&=+\frac{\alpha}{x}-\beta\,;\label{eqD1} \\
	\beta'&=-\frac{\beta}{x}-\alpha\,;\label{eqD2} \\
	-x\frac{T'}{T}&=8\pi G m T\left(\alpha\beta'-\alpha'\beta\right) \label{eqE2},
\end{align}
where now $'\equiv \dd/\dd x$, and (\ref{eqEE1twoF}) is automatically satisfied. The two Dirac equations can now be solved independently of the remaining Einstein equation:
\begin{align}
	\alpha(x)&=c_1e^{-x}+c_2e^{x};\\
	\beta(x)&=c_1\left(1+\frac{1}{x}\right)e^{-x}+c_2\left(1-\frac{1}{x}\right)e^{x},
\end{align}
where $c_1$ and $c_2$ are constants with dimensions $({\rm length})^{-1/2}$. In order to obtain a localized solution, we are forced to set $c_2=0$. Then, substituting these forms into (\ref{eqE2}), we obtain:
\begin{equation}
	\frac{T'}{T^2}=\frac{h_m}{x^3}\, e^{-2x},
\end{equation}
where we have defined the dimensionless constant $h_m=8\pi G c_1^2m$. This can be straightforwardly solved using integration by parts to give the following expression for $T$:
\begin{equation}
	T(x)=\frac{2x^2}{2c_3x^2+h_m\left[(1-2x)e^{-2x}-4x^2\mathrm{Ei}(-2x)\right]},
\end{equation}
where $\mathrm{Ei}(x)$ is the exponential integral. Imposing asymptotic flatness fixes $c_3=1$, leaving only a single integration constant $c_1$, which we shall relabel as $c_{m}$, since its value will depend on the fermion mass. The normalization condition (\ref{eqNormOv}), setting $A=1$, then becomes:
\begin{equation}
	2=4\pi\int_0^\infty x^2n_f\;\mathrm{d}x\label{eqNormRS},
\end{equation}
where the (radial) fermion density evaluates to:
\begin{equation}
	x^2n_f=\frac{4c_m^2m^{-1}(2x^2+2x+1)e^{-2x}}{2x^2+h_m\left[(1-2x)e^{-2x}-4x^2\mathrm{Ei}(-2x)\right]}.
\end{equation}

The normalization condition (\ref{eqNormRS}) is thus an implicit equation for $c_m$.  To check whether the integral on the right-hand side converges, consider the asymptotic behavior of the integrand:
\be
x^2n_f \sim \left\{ \begin{array}{lll} \displaystyle
\frac{1}{2\pi G m^2}\left(1+4x+\mathcal{O}\left(x^2\right)\right) & \quad & x \ll 1;\\
& & \\
\displaystyle \frac{4c_m^2}{m}\left(1+\frac{1}{x}+\mathcal{O}\left(\frac{1}{x^2}\right)\right)e^{-2x} & & x \gg 1.
\end{array} \right.
\ee
This confirms that the integrand is well-behaved at both $x=0$ and $x=\infty$. In addition, $(1-2x)e^{-2x}-4x^2\mathrm{Ei}(-2x)$ is strictly positive for all $x>0$, and so the fermion density is guaranteed to be positive at all radii. Hence we conclude that the solution is normalizable for any $m>0$.

\textit{Properties of the solution.}   
Overall, the solution can be written as:
\begin{eqnarray}
	\alpha(x)&=&c_{m}e^{-x};\\
	\beta(x)&=&c_{m}\left(1+\frac{1}{x}\right)e^{-x};\\
	T(x)&=& \left[ 1+h_m\left(\frac{1-2x}{2x^2 e^{2x}}-2\mathrm{Ei}(-2x)\right) \right]^{-1}. \quad
\end{eqnarray}
The metric field $T$ has the following asymptotic behavior:
\be
T\sim \left\{ \begin{array}{l} \displaystyle
\frac{2x^2}{h_m}\left(1+4x+\mathcal{O}(x^2) \right) \qquad \qquad \quad \;\;\, x \ll 1; \hspace{-10pt}\\
\\
\displaystyle 1-\frac{h_m}{2x^3 e^{2x}} \left(1-\frac{3}{2x}+\mathcal{O}\left(\frac{1}{x^2}\right)\right) \;\;\; x \gg 1.\hspace{-10pt}
\end{array} \right. \,\,\,\,
\ee
The fermion spinor field $\Psi_a$ decays exponentially at large $x$, but diverges at $x=0$; however, the zero in $T$ combines with this divergence in such a way as to maintain normalizability.  At large radii, the metric tends towards an asymptotically flat spacetime, but does so via an exponential decay; thus the total ADM mass of the solution, given by the coefficient of a $1/x$ term at large radius, is identically zero.

This can be confirmed by considering the Komar mass~\cite{WaldGR}, which here takes the following simple form:
\begin{align}
	M_{K}(x)=-\frac{x^2\sqrt{A}\,T'}{Gm T^2}=-\frac{h_m e^{-2x}}{Gmx}.
\end{align}
Since the ADM mass is defined as the large-$x$ limit of the Komar mass, we conclude that the total gravitational mass is zero, and thus the solution is gravitationally undetectable to an external observer:\ hence the name `stealth' solution.
The Komar mass is strictly negative for all radii, decreasing monotonically as $x\to 0$, with a divergence at $x=0$. This indicates that the object is gravitationally repulsive, with the force increasing in strength the nearer one approaches, ultimately becoming infinite at the center. This strongly suggests that the spacetime may contain a singularity at $x=0$, a conjecture that can be confirmed by evaluating the Ricci scalar:
\begin{align}
	\mathcal{R}(x)&=-8\pi G T^{\mu}_{\;\;\mu}=\frac{16\pi Gm^3}{x^2}T\left(\alpha^2-\beta^2\right)\notag\\
	&=\frac{-4m^2 h_m (2x
		+1)e^{-2x}}{2x^4-x^2h_m\left[(2x-1)e^{-2x}+4x^2\mathrm{Ei}(-2x)\right]}.
\end{align}
Expanding this at small $x$, we obtain:
\begin{align}
	\mathcal{R}(x) \approx -\frac{4m^2}{x^2} \left( 1+4x+\mathcal{O}\left(x^2\log x\right) \right),
\end{align}
which clearly diverges at $x=0$. This singularity is not accompanied, however, by an event horizon (a fact that can be confirmed by a null geodesic analysis); it is therefore a `naked' singularity, that in principle could be directly observed.

Further investigation of the solution's properties requires a numerical approach, details of which are provided in the supplemental material.  We find that, for each value of the fermion mass $m$, there exists a unique solution for the constant $c_m$.
\begin{figure}
	\includegraphics{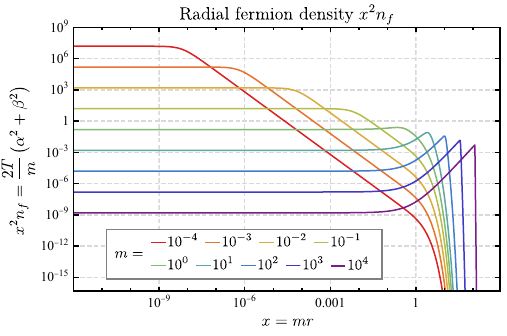}\\
	\includegraphics{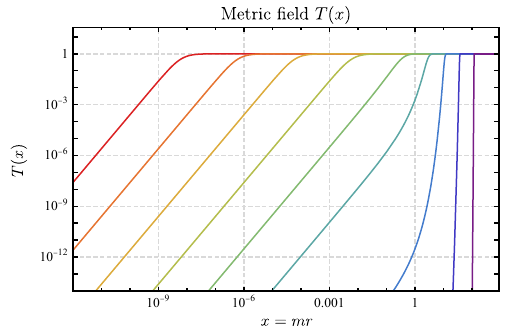}
	\caption{Upper panel: The radial fermion density, $x^2 n_f$, as a function of the normalized distance from the center of the object, $x$, for various values of the fermion mass $m$.  Note the crossover from a `solid ball' to a `hollow shell' structure as the fermion mass increases. Lower panel: The metric field $T$ as a function of the normalized distance from the center of the object, $x$, for the same values of the fermion mass $m$. Note that we have set $G=1$, without loss of generality.}
	\label{figRadialProfiles}
\end{figure}
The radial profiles for a selection of these solutions, with $m$ values ranging from $10^{-4}$ to $10^4$, are presented in Fig.~\ref{figRadialProfiles}.  There is a distinct change in behavior as the fermion mass is increased. For small $m$, the radial fermion density initially plateaus at small $x$ (until $x\approx 2Gm^2/\pi$), then decreases monotonically ($x^2n_f\sim1/x^2$) before ultimately decaying exponentially. As $m$ increases, the central value of $x^2n_f$ decreases, and the monotonically decreasing region is replaced by an increasing region ($x^2n_f\sim x^{3}$), before exponential decay occurs as before. The density profile thus transitions from a distribution that is peaked at $x=0$ (a `solid ball') to one that has a peak at non-zero radius (a `hollow shell'). 

The corresponding profiles for the metric field $T$ are also shown in Fig.~\ref{figRadialProfiles}. At small $m$, we find that $T$ initially exhibits a consistent increase ($T\sim x^2$) before asymptoting to $T=1$ (a flat metric). For larger $m$ this same increase still occurs, but is followed by a rapid exponential rise towards $T=1$.

This change in behavior can be partially understood by performing an asymptotic analysis in $m$, full details of which can be found in the supplemental material. At small $m$, we find that the radial fermion density and metric field $T$ asymptote to the following forms:
\begin{align}
&x^2n_f \sim \frac{4c_m^2m^{-1}}{2x^2+h_m};
&&T \sim \left[1+\frac{h_m}{2x^2}\right]^{-1},\label{eqSmallMforms}
\end{align}
with $c_m\sim m^{3/2}$. In contrast, at large $m$ we find:
\begin{align}
&x^2n_f \sim \frac{c_m^2x^3m^{-1}}{2x^3e^{2x}+h_m}; 
&&T \sim \left[1+\frac{h_m}{2x^3e^{2x}}\right]^{-1},\label{eqLargeMforms}
\end{align}
with $c_m\sim \left( m^2/G \right)^{1/8}\exp\left[ \left( 2Gm^2 \right)^{1/4}\right]$.

These expressions represent the extreme versions of the `solid ball' and `hollow shell' profiles shown in Fig.~\ref{figRadialProfiles}, to which the solution asymptotes at small and large $m$. Note that the localization mechanism is different in either case; for small $m$, the fermions are confined by a Newtonian-like $1/x^2$ term, whereas at large $m$ they are confined by an exponential decay.

\begin{figure}
	\includegraphics{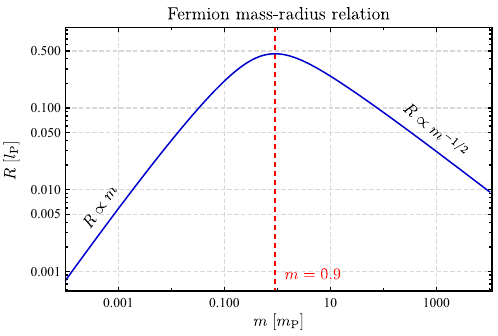}
	\caption{The radial extent of the fermionic matter distribution, $R$, as a function of the fermion mass, $m$. We have set $G=1$, allowing quantities to be written in Planck units. On the `solid ball' side of the diagram ($m \lesssim m_{\rm P}$) the object grows as the fermion mass increases; on the `hollow shell' side ($m \gtrsim m_{\rm P}$) it shrinks. Note that $R$ represents the physical radius of the solution, whereas the distributions in Fig.~\ref{figRadialProfiles} are plotted as a function of the rescaled radial co-ordinate $x=mr$.}
	\label{figMassRadius}
\end{figure}

The explicit dependence of the radial extent $R$ on the fermion mass is shown in Fig.~\ref{figMassRadius}, where we have defined $R$ as the average value of the (physical) co-ordinate radius $r$, weighted by the radial fermion density:
\begin{equation}
	R=4\pi\int_0^\infty rT\left(\alpha^2+\beta^2\right)\mathrm{d}r.
\end{equation}
The radial extent of all solutions is within one Planck radius, and shrinks to zero at both $m=0$ (as $R \propto m$) and $m=\infty$ (as $R \propto m^{-1/2}$), in agreement with an analytical asymptotic analysis (see supplemental material for details).  The transition in behavior from `solid ball' to `hollow sphere' occurs at the order of one Planck mass.

These results demonstrate that the stealth solution remains highly localized, irrespective of the value of the fermion mass. This is counter-intuitive:\ one might have expected the weak gravitational attraction that exists between low-mass fermions to result in a bound state that extends over a large volume (as seen, indeed, in the original non-singular FSY solutions). For the stealth solution, however, the self-consistently determined properties of the naked singularity alter the situation significantly, preventing the system from ever reaching a classical limit.

\textit{Discussion.}  We have demonstrated that the Einstein--Dirac equation system admits a `stealth' solution with the following properties:\ a fermionic wavefunction that is exponentially localized and normalizable; a spacetime containing a naked singularity at the origin; and zero ADM mass. How should we interpret the existence of such a solution, and how physical is it?

Perhaps the most surprising feature of the solution is its vanishing ADM mass, combined with the observation --- see Fig.~\ref{figMassRadius} --- that the radial extent of the object tends to zero both as $m \to 0$ and as $m \to \infty$.  This implies that arbitrarily large masses could be `hidden' in arbitrarily small spatial volumes. This behavior is at odds with our usual interpretation of how matter and gravity interact, suggesting caution when applying intuition from classical gravity to the quantum regime.

Although the fermion mass $m$ can take any positive value, the fermion energy $\omega$ is identically zero, i.e.\ the stealth solution `lives' precisely mid-way between the particle continuum ($\omega \geqslant m$) and the antiparticle continuum ($\omega \leqslant -m$). The ADM mass and the fermion energy are thus completely decoupled from the fermion mass.  Curiously, $\omega=0$ here does {\it not\/} imply that the particle and antiparticle components of the wavefunction, $\alpha$ and $\beta$, have the same functional form; indeed, one of these diverges at the origin while the other does not.  This is in contrast to the case of Majorana zero modes, which are `pinned' to the center of the spectral gap precisely because they are their own antiparticles~\cite{Sarma2015}.

The existence of a solution with such properties is likely related to the presence of the naked singularity. In a classical context, naked singularities are associated with unphysicality of the matter sector, and indeed we find that the stealth solution violates the weak, strong, and dominant energy conditions (see supplemental material for details). It is not clear, however, whether one would expect such conditions to hold in the context of quantum-mechanical matter~\cite{Kontou2020}. Indeed, since our fermion wavefunction is a valid solution to the Dirac equation, and moreover is normalizable, it would appear to be a well-defined quantum state. Exploring this question further, especially in the context of these stealth solutions, would be a useful subject for future research. We note in passing that a similar result concerning normalizability and naked singularities has been reported for Dirac fields in a Reissner--Nordstr{\"o}m background~\cite{Batic2011NakedSingularitiesED}.

Finally, we should emphasize that the Einstein--Dirac system studied here is not a full theory of quantum gravity, but only a semi-classical approximation. It is worth noting, however, that the crossover from `solid ball' to `hollow shell' solutions occurs at $m \sim m_{\rm P}$, even though our theoretical approach neglects much of the expected physics of quantum gravity (e.g.\ radiative corrections due to the graviton field).

\textit{Acknowledgments.} P.E.D.L. acknowledges funding from a St Leonards scholarship from the University of St Andrews and from UKRI under EPSRC Grant No.~EP/R513337/1.

\bibliography{references}

\begin{thebibliography}{23}%
\makeatletter
\providecommand \@ifxundefined [1]{%
 \@ifx{#1\undefined}
}%
\providecommand \@ifnum [1]{%
 \ifnum #1\expandafter \@firstoftwo
 \else \expandafter \@secondoftwo
 \fi
}%
\providecommand \@ifx [1]{%
 \ifx #1\expandafter \@firstoftwo
 \else \expandafter \@secondoftwo
 \fi
}%
\providecommand \natexlab [1]{#1}%
\providecommand \enquote  [1]{``#1''}%
\providecommand \bibnamefont  [1]{#1}%
\providecommand \bibfnamefont [1]{#1}%
\providecommand \citenamefont [1]{#1}%
\providecommand \href@noop [0]{\@secondoftwo}%
\providecommand \href [0]{\begingroup \@sanitize@url \@href}%
\providecommand \@href[1]{\@@startlink{#1}\@@href}%
\providecommand \@@href[1]{\endgroup#1\@@endlink}%
\providecommand \@sanitize@url [0]{\catcode `\\12\catcode `\$12\catcode
  `\&12\catcode `\#12\catcode `\^12\catcode `\_12\catcode `\%12\relax}%
\providecommand \@@startlink[1]{}%
\providecommand \@@endlink[0]{}%
\providecommand \url  [0]{\begingroup\@sanitize@url \@url }%
\providecommand \@url [1]{\endgroup\@href {#1}{\urlprefix }}%
\providecommand \urlprefix  [0]{URL }%
\providecommand \Eprint [0]{\href }%
\providecommand \doibase [0]{https://doi.org/}%
\providecommand \selectlanguage [0]{\@gobble}%
\providecommand \bibinfo  [0]{\@secondoftwo}%
\providecommand \bibfield  [0]{\@secondoftwo}%
\providecommand \translation [1]{[#1]}%
\providecommand \BibitemOpen [0]{}%
\providecommand \bibitemStop [0]{}%
\providecommand \bibitemNoStop [0]{.\EOS\space}%
\providecommand \EOS [0]{\spacefactor3000\relax}%
\providecommand \BibitemShut  [1]{\csname bibitem#1\endcsname}%
\let\auto@bib@innerbib\@empty
\bibitem [{\citenamefont {Joshi}(2007)}]{JoshiSing}%
  \BibitemOpen
  \bibfield  {author} {\bibinfo {author} {\bibfnamefont {P.~S.}\ \bibnamefont
  {Joshi}},\ }\href@noop {} {\emph {\bibinfo {title} {Gravitational collapse
  and space-time singularities}}}\ (\bibinfo  {publisher} {Cambridge University
  Press},\ \bibinfo {year} {2007})\BibitemShut {NoStop}%
\bibitem [{\citenamefont {{The Event Horizon Telescope Collaboration {\it et
  al.\/}}}(2019)}]{EHTcollab}%
  \BibitemOpen
  \bibfield  {author} {\bibinfo {author} {\bibnamefont {{The Event Horizon
  Telescope Collaboration {\it et al.\/}}}},\ }\bibfield  {title} {\bibinfo
  {title} {{First M87 Event Horizon Telescope Results. I. The Shadow of the
  Supermassive Black Hole}},\ }\href {https://doi.org/10.3847/2041-8213/ab0ec7}
  {\bibfield  {journal} {\bibinfo  {journal} {Astrophys. J. Lett.}\ }\textbf
  {\bibinfo {volume} {875}},\ \bibinfo {pages} {L1} (\bibinfo {year}
  {2019})}\BibitemShut {NoStop}%
\bibitem [{\citenamefont {{B. P. Abbott {\it et al.}}}(2016)}]{Abbott2016}%
  \BibitemOpen
  \bibfield  {author} {\bibinfo {author} {\bibnamefont {{B. P. Abbott {\it et
  al.}}}} (\bibinfo {collaboration} {LIGO Scientific Collaboration and Virgo
  Collaboration}),\ }\bibfield  {title} {\bibinfo {title} {Binary black hole
  mergers in the first advanced {LIGO} observing run},\ }\href
  {https://doi.org/10.1103/PhysRevX.6.041015} {\bibfield  {journal} {\bibinfo
  {journal} {Phys. Rev. X}\ }\textbf {\bibinfo {volume} {6}},\ \bibinfo {pages}
  {041015} (\bibinfo {year} {2016})}\BibitemShut {NoStop}%
\bibitem [{\citenamefont {Bondi}(1957)}]{Bondi1957}%
  \BibitemOpen
  \bibfield  {author} {\bibinfo {author} {\bibfnamefont {H.}~\bibnamefont
  {Bondi}},\ }\bibfield  {title} {\bibinfo {title} {Negative mass in general
  relativity},\ }\href {https://doi.org/10.1103/RevModPhys.29.423} {\bibfield
  {journal} {\bibinfo  {journal} {Rev. Mod. Phys.}\ }\textbf {\bibinfo {volume}
  {29}},\ \bibinfo {pages} {423} (\bibinfo {year} {1957})}\BibitemShut
  {NoStop}%
\bibitem [{\citenamefont {Bellet\^{e}te}\ and\ \citenamefont
  {Paranjape}(2013)}]{Belletete2013}%
  \BibitemOpen
  \bibfield  {author} {\bibinfo {author} {\bibfnamefont {J.}~\bibnamefont
  {Bellet\^{e}te}}\ and\ \bibinfo {author} {\bibfnamefont {M.~B.}\ \bibnamefont
  {Paranjape}},\ }\bibfield  {title} {\bibinfo {title} {On negative mass},\
  }\href {https://doi.org/10.1142/S0218271813410174} {\bibfield  {journal}
  {\bibinfo  {journal} {Int. J. Mod. Phys. D}\ }\textbf {\bibinfo {volume}
  {22}},\ \bibinfo {pages} {1341017} (\bibinfo {year} {2013})}\BibitemShut
  {NoStop}%
\bibitem [{\citenamefont {Joshi}\ and\ \citenamefont
  {Bhattacharyya}(2025)}]{joshi2024}%
  \BibitemOpen
  \bibfield  {author} {\bibinfo {author} {\bibfnamefont {P.~S.}\ \bibnamefont
  {Joshi}}\ and\ \bibinfo {author} {\bibfnamefont {S.}~\bibnamefont
  {Bhattacharyya}},\ }\bibfield  {title} {\bibinfo {title} {Primordial naked
  singularities},\ }\href {https://doi.org/10.1088/1475-7516/2025/01/034}
  {\bibfield  {journal} {\bibinfo  {journal} {J. Cosmol. Astropart. Phys.}\
  }\textbf {\bibinfo {volume} {2025}}\bibinfo  {number} { (01)},\ \bibinfo
  {pages} {034}}\BibitemShut {NoStop}%
\bibitem [{\citenamefont {Klinkhamer}\ and\ \citenamefont
  {Queiruga}(2018)}]{Klinkhamer2018stealth}%
  \BibitemOpen
\bibfield  {number} {  }\bibfield  {author} {\bibinfo {author} {\bibfnamefont
  {F.~R.}\ \bibnamefont {Klinkhamer}}\ and\ \bibinfo {author} {\bibfnamefont
  {J.~M.}\ \bibnamefont {Queiruga}},\ }\bibfield  {title} {\bibinfo {title} {A
  stealth defect of spacetime},\ }\href
  {https://doi.org/10.1142/S0217732318501274} {\bibfield  {journal} {\bibinfo
  {journal} {Mod. Phys. Lett. A}\ }\textbf {\bibinfo {volume} {33}},\ \bibinfo
  {pages} {1850127} (\bibinfo {year} {2018})}\BibitemShut {NoStop}%
\bibitem [{\citenamefont {Klinkhamer}\ and\ \citenamefont
  {Wang}(2019)}]{Klinkhamer2019lensing}%
  \BibitemOpen
  \bibfield  {author} {\bibinfo {author} {\bibfnamefont {F.~R.}\ \bibnamefont
  {Klinkhamer}}\ and\ \bibinfo {author} {\bibfnamefont {Z.~L.}\ \bibnamefont
  {Wang}},\ }\bibfield  {title} {\bibinfo {title} {Lensing and imaging by a
  stealth defect of spacetime},\ }\href
  {https://doi.org/10.1142/S0217732319500263} {\bibfield  {journal} {\bibinfo
  {journal} {Mod. Phys. Lett. A}\ }\textbf {\bibinfo {volume} {34}},\ \bibinfo
  {pages} {1950026} (\bibinfo {year} {2019})}\BibitemShut {NoStop}%
\bibitem [{\citenamefont {Klinkhamer}(2019)}]{Klinkhamer2019soliton}%
  \BibitemOpen
  \bibfield  {author} {\bibinfo {author} {\bibfnamefont {F.~R.}\ \bibnamefont
  {Klinkhamer}},\ }\bibfield  {title} {\bibinfo {title} {On a soliton-type
  spacetime defect},\ }\href {https://doi.org/10.1088/1742-6596/1275/1/012012}
  {\bibfield  {journal} {\bibinfo  {journal} {J. Phys. Conf. Ser.}\ }\textbf
  {\bibinfo {volume} {1275}},\ \bibinfo {pages} {012012} (\bibinfo {year}
  {2019})}\BibitemShut {NoStop}%
\bibitem [{\citenamefont {Finster}\ \emph
  {et~al.}(1999{\natexlab{a}})\citenamefont {Finster}, \citenamefont
  {Smoller},\ and\ \citenamefont {Yau}}]{FSY1999original}%
  \BibitemOpen
  \bibfield  {author} {\bibinfo {author} {\bibfnamefont {F.}~\bibnamefont
  {Finster}}, \bibinfo {author} {\bibfnamefont {J.}~\bibnamefont {Smoller}},\
  and\ \bibinfo {author} {\bibfnamefont {S.-T.}\ \bibnamefont {Yau}},\
  }\bibfield  {title} {\bibinfo {title} {Particlelike solutions of the
  {E}instein-{D}irac equations},\ }\href
  {https://doi.org/10.1103/PhysRevD.59.104020} {\bibfield  {journal} {\bibinfo
  {journal} {Phys. Rev. D}\ }\textbf {\bibinfo {volume} {59}},\ \bibinfo
  {pages} {104020} (\bibinfo {year} {1999}{\natexlab{a}})}\BibitemShut
  {NoStop}%
\bibitem [{\citenamefont {Finster}\ \emph
  {et~al.}(1999{\natexlab{b}})\citenamefont {Finster}, \citenamefont
  {Smoller},\ and\ \citenamefont {Yau}}]{FSY1999maxwell}%
  \BibitemOpen
  \bibfield  {author} {\bibinfo {author} {\bibfnamefont {F.}~\bibnamefont
  {Finster}}, \bibinfo {author} {\bibfnamefont {J.}~\bibnamefont {Smoller}},\
  and\ \bibinfo {author} {\bibfnamefont {S.-T.}\ \bibnamefont {Yau}},\
  }\bibfield  {title} {\bibinfo {title} {Particle-like solutions of the
  {Einstein}--{D}irac--{M}axwell equations},\ }\href
  {https://doi.org/10.1016/S0375-9601(99)00457-0} {\bibfield  {journal}
  {\bibinfo  {journal} {Phys. Lett. A}\ }\textbf {\bibinfo {volume} {259}},\
  \bibinfo {pages} {431} (\bibinfo {year} {1999}{\natexlab{b}})}\BibitemShut
  {NoStop}%
\bibitem [{\citenamefont {Finster}\ \emph {et~al.}(2000)\citenamefont
  {Finster}, \citenamefont {Smoller},\ and\ \citenamefont
  {Yau}}]{FSY2000nonAbelianBound}%
  \BibitemOpen
  \bibfield  {author} {\bibinfo {author} {\bibfnamefont {F.}~\bibnamefont
  {Finster}}, \bibinfo {author} {\bibfnamefont {J.}~\bibnamefont {Smoller}},\
  and\ \bibinfo {author} {\bibfnamefont {S.-T.}\ \bibnamefont {Yau}},\
  }\bibfield  {title} {\bibinfo {title} {The interaction of {D}irac particles
  with non-{A}belian gauge fields and gravity--bound states},\ }\href
  {https://doi.org/10.1016/S0550-3213(00)00370-9} {\bibfield  {journal}
  {\bibinfo  {journal} {Nucl. Phys. B}\ }\textbf {\bibinfo {volume} {584}},\
  \bibinfo {pages} {387} (\bibinfo {year} {2000})}\BibitemShut {NoStop}%
\bibitem [{\citenamefont {Leith}\ \emph {et~al.}(2023)\citenamefont {Leith},
  \citenamefont {Leggat}, \citenamefont {Hooley}, \citenamefont {Horne},\ and\
  \citenamefont {Dritschel}}]{Leith2023Higgs}%
  \BibitemOpen
  \bibfield  {author} {\bibinfo {author} {\bibfnamefont {P.~E.}\ \bibnamefont
  {Leith}}, \bibinfo {author} {\bibfnamefont {A.~D.}\ \bibnamefont {Leggat}},
  \bibinfo {author} {\bibfnamefont {C.~A.}\ \bibnamefont {Hooley}}, \bibinfo
  {author} {\bibfnamefont {K.}~\bibnamefont {Horne}},\ and\ \bibinfo {author}
  {\bibfnamefont {D.~G.}\ \bibnamefont {Dritschel}},\ }\bibfield  {title}
  {\bibinfo {title} {Gravitationally localized states of two neutral fermions
  interacting with a higgs field},\ }\href
  {https://doi.org/10.1103/PhysRevD.107.106020} {\bibfield  {journal} {\bibinfo
   {journal} {Phys. Rev. D}\ }\textbf {\bibinfo {volume} {107}},\ \bibinfo
  {pages} {106020} (\bibinfo {year} {2023})}\BibitemShut {NoStop}%
\bibitem [{\citenamefont {Leith}\ \emph {et~al.}(2020)\citenamefont {Leith},
  \citenamefont {Hooley}, \citenamefont {Horne},\ and\ \citenamefont
  {Dritschel}}]{Leith2020fermionTrapping}%
  \BibitemOpen
  \bibfield  {author} {\bibinfo {author} {\bibfnamefont {P.~E.~D.}\
  \bibnamefont {Leith}}, \bibinfo {author} {\bibfnamefont {C.~A.}\ \bibnamefont
  {Hooley}}, \bibinfo {author} {\bibfnamefont {K.}~\bibnamefont {Horne}},\ and\
  \bibinfo {author} {\bibfnamefont {D.~G.}\ \bibnamefont {Dritschel}},\
  }\bibfield  {title} {\bibinfo {title} {Fermion self-trapping in the optical
  geometry of {E}instein-{D}irac solitons},\ }\href
  {https://doi.org/10.1103/PhysRevD.101.106012} {\bibfield  {journal} {\bibinfo
   {journal} {Phys. Rev. D}\ }\textbf {\bibinfo {volume} {101}},\ \bibinfo
  {pages} {106012} (\bibinfo {year} {2020})}\BibitemShut {NoStop}%
\bibitem [{\citenamefont {Leith}\ \emph {et~al.}(2021)\citenamefont {Leith},
  \citenamefont {Hooley}, \citenamefont {Horne},\ and\ \citenamefont
  {Dritschel}}]{Leith2021excited}%
  \BibitemOpen
  \bibfield  {author} {\bibinfo {author} {\bibfnamefont {P.~E.~D.}\
  \bibnamefont {Leith}}, \bibinfo {author} {\bibfnamefont {C.~A.}\ \bibnamefont
  {Hooley}}, \bibinfo {author} {\bibfnamefont {K.}~\bibnamefont {Horne}},\ and\
  \bibinfo {author} {\bibfnamefont {D.~G.}\ \bibnamefont {Dritschel}},\
  }\bibfield  {title} {\bibinfo {title} {Nonlinear effects in the excited
  states of many-fermion {E}instein-{D}irac solitons},\ }\href
  {https://doi.org/10.1103/PhysRevD.104.046024} {\bibfield  {journal} {\bibinfo
   {journal} {Phys. Rev. D}\ }\textbf {\bibinfo {volume} {104}},\ \bibinfo
  {pages} {046024} (\bibinfo {year} {2021})}\BibitemShut {NoStop}%
\bibitem [{\citenamefont {Herdeiro}\ \emph {et~al.}(2019)\citenamefont
  {Herdeiro}, \citenamefont {Perapechka}, \citenamefont {Radu},\ and\
  \citenamefont {Shnir}}]{Herdeiro2019bosonDiracProcaSpinning}%
  \BibitemOpen
  \bibfield  {author} {\bibinfo {author} {\bibfnamefont {C.}~\bibnamefont
  {Herdeiro}}, \bibinfo {author} {\bibfnamefont {P.}~\bibnamefont
  {Perapechka}}, \bibinfo {author} {\bibfnamefont {E.}~\bibnamefont {Radu}},\
  and\ \bibinfo {author} {\bibfnamefont {Y.}~\bibnamefont {Shnir}},\ }\bibfield
   {title} {\bibinfo {title} {Asymptotically flat spinning scalar, {D}irac and
  {P}roca stars},\ }\href {https://doi.org/10.1016/j.physletb.2019.134845}
  {\bibfield  {journal} {\bibinfo  {journal} {Phys. Lett. B}\ }\textbf
  {\bibinfo {volume} {797}},\ \bibinfo {pages} {134845} (\bibinfo {year}
  {2019})}\BibitemShut {NoStop}%
\bibitem [{\citenamefont {Bakucz~Can{\'a}rio}\ \emph
  {et~al.}(2020)\citenamefont {Bakucz~Can{\'a}rio}, \citenamefont {Lloyd},
  \citenamefont {Horne},\ and\ \citenamefont {Hooley}}]{Bakucz2020powerlaw}%
  \BibitemOpen
  \bibfield  {author} {\bibinfo {author} {\bibfnamefont {D.}~\bibnamefont
  {Bakucz~Can{\'a}rio}}, \bibinfo {author} {\bibfnamefont {S.}~\bibnamefont
  {Lloyd}}, \bibinfo {author} {\bibfnamefont {K.}~\bibnamefont {Horne}},\ and\
  \bibinfo {author} {\bibfnamefont {C.~A.}\ \bibnamefont {Hooley}},\ }\bibfield
   {title} {\bibinfo {title} {Infinite-redshift localized states of {D}irac
  fermions under {E}insteinian gravity},\ }\href
  {https://doi.org/10.1103/PhysRevD.102.084049} {\bibfield  {journal} {\bibinfo
   {journal} {Phys. Rev. D}\ }\textbf {\bibinfo {volume} {102}},\ \bibinfo
  {pages} {084049} (\bibinfo {year} {2020})}\BibitemShut {NoStop}%
\bibitem [{\citenamefont {Bl{\'a}zquez-Salcedo}\ and\ \citenamefont
  {Knoll}(2020)}]{Blazquez2020ansatz}%
  \BibitemOpen
  \bibfield  {author} {\bibinfo {author} {\bibfnamefont {J.~L.}\ \bibnamefont
  {Bl{\'a}zquez-Salcedo}}\ and\ \bibinfo {author} {\bibfnamefont
  {C.}~\bibnamefont {Knoll}},\ }\bibfield  {title} {\bibinfo {title}
  {Constructing spherically symmetric {E}instein--{D}irac systems with multiple
  spinors: {A}nsatz, wormholes and other analytical solutions},\ }\href
  {https://doi.org/10.1140/epjc/s10052-020-7706-3} {\bibfield  {journal}
  {\bibinfo  {journal} {Eur.~Phys.~J.~C.}\ }\textbf {\bibinfo {volume} {80}},\
  \bibinfo {pages} {174} (\bibinfo {year} {2020})}\BibitemShut {NoStop}%
\bibitem [{\citenamefont {Wald}(2010)}]{WaldGR}%
  \BibitemOpen
  \bibfield  {author} {\bibinfo {author} {\bibfnamefont {R.~M.}\ \bibnamefont
  {Wald}},\ }\href@noop {} {\emph {\bibinfo {title} {General relativity}}}\
  (\bibinfo  {publisher} {University of Chicago press},\ \bibinfo {year}
  {2010})\BibitemShut {NoStop}%
\bibitem [{\citenamefont {Sarma}\ \emph {et~al.}(2015)\citenamefont {Sarma},
  \citenamefont {Freedman},\ and\ \citenamefont {Nayak}}]{Sarma2015}%
  \BibitemOpen
  \bibfield  {author} {\bibinfo {author} {\bibfnamefont {S.}~\bibnamefont
  {Sarma}}, \bibinfo {author} {\bibfnamefont {M.}~\bibnamefont {Freedman}},\
  and\ \bibinfo {author} {\bibfnamefont {C.}~\bibnamefont {Nayak}},\ }\bibfield
   {title} {\bibinfo {title} {Majorana zero modes and topological quantum
  computation},\ }\href {https://doi.org/10.1038/npjqi.2015.1} {\bibfield
  {journal} {\bibinfo  {journal} {npj Quantum Inf.}\ }\textbf {\bibinfo
  {volume} {1}},\ \bibinfo {pages} {15001} (\bibinfo {year}
  {2015})}\BibitemShut {NoStop}%
\bibitem [{\citenamefont {Kontou}\ and\ \citenamefont
  {Sanders}(2020)}]{Kontou2020}%
  \BibitemOpen
  \bibfield  {author} {\bibinfo {author} {\bibfnamefont {E.-A.}\ \bibnamefont
  {Kontou}}\ and\ \bibinfo {author} {\bibfnamefont {K.}~\bibnamefont
  {Sanders}},\ }\bibfield  {title} {\bibinfo {title} {Energy conditions in
  general relativity and quantum field theory},\ }\href
  {https://doi.org/10.1088/1361-6382/ab8fcf} {\bibfield  {journal} {\bibinfo
  {journal} {Class. Quant. Grav.}\ }\textbf {\bibinfo {volume} {37}},\ \bibinfo
  {pages} {193001} (\bibinfo {year} {2020})}\BibitemShut {NoStop}%
\bibitem [{\citenamefont {Batic}\ \emph {et~al.}(2011)\citenamefont {Batic},
  \citenamefont {Chin},\ and\ \citenamefont
  {Nowakowski}}]{Batic2011NakedSingularitiesED}%
  \BibitemOpen
  \bibfield  {author} {\bibinfo {author} {\bibfnamefont {D.}~\bibnamefont
  {Batic}}, \bibinfo {author} {\bibfnamefont {D.}~\bibnamefont {Chin}},\ and\
  \bibinfo {author} {\bibfnamefont {M.}~\bibnamefont {Nowakowski}},\ }\bibfield
   {title} {\bibinfo {title} {The repulsive nature of naked singularities from
  the point of view of quantum mechanics},\ }\href
  {https://doi.org/10.1140/epjc/s10052-011-1624-3} {\bibfield  {journal}
  {\bibinfo  {journal} {Eur. Phys. J. C}\ }\textbf {\bibinfo {volume} {71}},\
  \bibinfo {pages} {1624} (\bibinfo {year} {2011})}\BibitemShut {NoStop}%
\bibitem [{\citenamefont {Finster}\ \emph
  {et~al.}(1999{\natexlab{c}})\citenamefont {Finster}, \citenamefont
  {Smoller},\ and\ \citenamefont {Yau}}]{FSY1999bhEDM}%
  \BibitemOpen
  \bibfield  {author} {\bibinfo {author} {\bibfnamefont {F.}~\bibnamefont
  {Finster}}, \bibinfo {author} {\bibfnamefont {J.}~\bibnamefont {Smoller}},\
  and\ \bibinfo {author} {\bibfnamefont {S.-T.}\ \bibnamefont {Yau}},\
  }\bibfield  {title} {\bibinfo {title} {Non-existence of black hole solutions
  for a spherically symmetric, static {E}instein--{D}irac--{M}axwell system},\
  }\href {https://doi.org/10.1007/s002200050675} {\bibfield  {journal}
  {\bibinfo  {journal} {Commun. Math. Phys}\ }\textbf {\bibinfo {volume}
  {205}},\ \bibinfo {pages} {249} (\bibinfo {year}
  {1999}{\natexlab{c}})}\BibitemShut {NoStop}%
\end{thebibliography}%

\clearpage
\onecolumngrid
\begin{center}
{\large \bf{Supplemental material for ```Stealth' singularities from self-gravitating fermions''}}
\end{center}
\vspace{15pt}

\setcounter{equation}{0}
\setcounter{figure}{0}
\setcounter{page}{1}
\def\theequation{S.\arabic{equation}}
\def\thefigure{S.\arabic{figure}}
\def\thepage{S\arabic{page}}

\twocolumngrid

\section{Numerical method}

Although the stealth solution presented in the main text can be written analytically, a numerical element is nonetheless required to obtain an explicit solution for a given value of $m$, owing to the integral nature of the normalization condition that defines the value of $c_m$. The procedure for generating solutions can be summarized as follows. For a chosen value of $m$, we first evaluate the normalisation integral (\ref{eqNormRS}) numerically using \textit{Mathematica}'s built-in differential equation solver, NDSolve, initially choosing an arbitrary value of $c_m=1$. Based on the magnitude of the result, we then re-perform the integration with $c_m$ either halved or doubled, iterating this until the region in which the integral evaluates to one has been identified. A simple binary chop is then performed (re-evaluating the integral at each step) to determine the precise value of $c_m$ for which the solution is correctly normalized. Having obtained the correct value of $c_m$, this can be straightforwardly substituted into the analytic expressions for $\alpha$, $\beta$ and $T$. 

\section{The many-fermion system}

The two-fermion system originally studied by FSY has since been extended to incorporate an arbitrarily large (even) number of fermions $N$, by arranging them in a single filled shell with angular momentum $j=(N-1)/2$. A derivation of this system is provided in~\cite{FSY1999bhEDM} and analyses of the associated localized solutions can be found in~\cite{Leith2020fermionTrapping} and~\cite{Leith2021excited}. Here, we show there exists also a many-fermion equivalent of the stealth solution, which can be written in a simple analytic form provided the number of fermions is even.

Generalizing the Einstein--Dirac system to many fermions is straightforward, with the equations of motion taking the following form (with $A=1$ and $\omega=0$):
\begin{align}
	\alpha'&=+\frac{N\alpha}{2x}-\beta;\label{eqD1many} \\
	\beta'&=-\frac{N\beta}{2x}-\alpha;\label{eqD2many} \\
	-x\frac{T'}{T}&=4\pi G m N T\left(\alpha\beta'-\alpha'\beta\right) \label{eqE2many}.
\end{align}
Note that here, as in the main text, we are restricting to positive parity solutions, with the acknowledgment that these are structurally equivalent to those of negative parity, as shown in the proceeding section. As in the two-fermion case, the two Dirac equations can be solved independently of the Einstein equation, with $\alpha$ and $\beta$ decoupling as follows:
\begin{align}
    \alpha''&=\left(1-\frac{N}{2x^2}+\frac{N^2}{4x^2}\right)\alpha; \\
    \beta''&=\left(1+\frac{N}{2x^2}+\frac{N^2}{4x^2}\right)\beta.
\end{align}
The general solution to these can be written in terms of modified Bessel functions:
\begin{align}
\alpha(x)&=\sqrt{x}\left(c_1 K_j(x)+c_2 I_j(x)\right);\\
\beta(x)&=\sqrt{x}\left(c_1 K_{j+1}(x)+c_2 I_{j+1}(x)\right).
\end{align}
Restricting to normalizable solutions forces us to take $c_2=0$, and, substituting these expressions into the Einstein equation, we then obtain:
\begin{align}
&4 \frac{T'}{T^2}=Nh_m \Bigl(K_j(x)\left[K_j(x)+K_{j+2}(x)\right] \notag\\
&\hspace{65pt}-K_{j+1}(x)\left[K_{j-1}(x)+K_{j+1}(x)\right]\Bigl).
\end{align}
For the case of an even number of fermions (the physically acceptable case), we can utilize the properties of the half-integer Bessel functions, along with repeated integration by parts, to obtain an expression for $T$ that takes the following general form:
\begin{align}
&T(x)=\left[1+h_m\left(\frac{p_N(x)}{2x^Ne^{2x}}-(-1)^{\frac{N}{2}}N\mathrm{Ei}(-2x)\right)\right]^{-1},
\end{align}
where the polynomial function $p_N$ differs depending on the value of $N$, with the first few being:
\begin{align}
    p_2(x)&=1-2x;\\
    p_4(x)&=3+6x-2x^2+4x^3;\\
    p_6(x)&=45 + 90 x + 63 x^2 + 6 x^3 + 3 x^4 - 6 x^5;\\
    p_8(x)&=1575 + 3150 x + 2670 x^2 + 1140 x^3 \notag \\
    &\hspace{50pt}+ 210 x^4 + 20 x^5 - 4 x^6 + 8 x^7.
\end{align}

We find that the resulting numerical solutions, which can be generated via the same method outlined above, exhibit a similar behavior to their two-fermion counterparts, with again a transition from `solid ball' to `hollow shell' as $m$ increases. The asymptotic relationship between the fermion mass and the radial extent is also found to be identical, although the fermion mass at which the radial extent peaks increases monotonically with increasing fermion number.

\section{Negative parity (and negative fermion mass) solutions for {\boldmath $N=2$}}

The stealth solutions presented in the main text were constructed under the assumption that the fermions have positive parity. In their analysis of the full Einstein--Dirac system, FSY demonstrated that negative parity solutions also exist, which have distinct properties from their positive-parity counterparts (most notably an additional node in the fermion wavefunction). We shall show here that such distinct solutions do not exist in the context of the stealth solution, with negative parity states being physically indistinct from those of positive parity. We also show that states with negative fermion mass exist, and are equivalent to those with positive fermion mass.

Negative parity solutions can be generated by modifying the fermion ansatz (\ref{eq2spinorAns2F}) to take the following form:
\begin{equation}
    \Psi_a(r,t)=\frac{\sqrt{T(r)}}{r}\binom{i\sigma^r\beta(r)e_a}{\alpha(r)e_a}e^{-i\omega t},
\end{equation}
i.e.\ the transformation of the positive-parity ansatz under $\gamma^5$. Taking this into account, a generalized form for the equations of motion, valid for both signs of parity, can be written as:
\begin{align}
	\alpha'&=+\frac{\alpha}{r}-\sigma m\beta; \\
	\beta'&=-\frac{\beta}{r}-\sigma m\alpha; \\
	-r\frac{T'}{T}&=8\pi G T\left(\alpha\beta'-\alpha'\beta\right),
\end{align}
where $'\equiv \mathrm{d}/\mathrm{d}r$, and $\sigma=\pm 1$ denotes positive or negative parity. Note that we have explicitly reintroduced the physical radius $r$, to avoid confusion when discussing negative fermion mass. When written in this form, it becomes clear that there is a symmetry between the sign of the fermion mass and the sign of the parity; solutions with positive parity and fermion mass $m$ are equivalent to those with negative parity and fermion mass $-m$. This symmetry also exists in the full Einstein--Dirac system, restricting the number of structurally distinct families of (ground-state) solutions to two.

For the stealth solution, it is straightforward to solve the generalized equations of motion to obtain the following expressions, valid for both possible signs of parity and fermion mass:
\begin{align}
	\alpha(r)&=c_me^{-|m|r};\\
	\beta(r)&=c_m\left(1+\frac{1}{|m|r}\right)e^{-|m|r};\\
    T(r)&=\left[1+h_m\left(\frac{1-2|m|r}{2|m|r^2 e^{2|m|r}}-2\mathrm{Ei}(-2|m|r)\right) \right]^{-1}.
\end{align}
Note that the parity $\sigma$ has dropped out entirely, and that the only change from the equations presented in the main text is the replacement of $m$ with $|m|$ (a consequence of restricting to normalizable solutions). This therefore implies that, for the stealth solution, there is no structural distinction between positive parity and negative parity solutions, or, equivalently, positive (fermion) mass and negative (fermion) mass solutions.

Although interesting, this feature is perhaps not surprising, since the stealth solution is generated by setting $\omega=0$, and one might therefore expect an additional symmetry to arise in the system when compared to the full Einstein--Dirac case.

\section{Asymptotic analysis in $\boldsymbol{m}$}

A particularly interesting feature of the stealth solution is the numerically observed transition from `solid ball' to `hollow sphere' as the fermion mass increases. To investigate this, we shall here perform an asymptotic analysis on the equations of motion, for both small and large values of $m$, allowing us to derive the asymptotic forms of $n_f$ and $T$ stated in (\ref{eqSmallMforms}) and (\ref{eqLargeMforms}), as well as the asymptotic mass-radius relations shown in Fig.~\ref{figMassRadius}.

For ease of reference, we first reproduce the general (two-fermion) expressions for the fermion density $n_f$ and metric field $T$:
\begin{align}
n_f&=\frac{4c_m^2 m^{-1}(2x^2+2x+1)e^{-2x}}{2x^2-h_m\left[(2x-1)e^{-2x}+4x^2\mathrm{Ei}(-2x)\right]};\label{eqSuppNfMaster}\\
T&= \left[ 1+h_m\left(\frac{1-2x}{2x^2 e^{2x}}-2\mathrm{Ei}(-2x)\right) \right]^{-1}\label{eqSuppTMaster},
\end{align}
where recall $x=mr$, $h_m=8\pi Gc_m^2m$, and the value of $c_m$ is defined by imposing the normalization condition: \begin{equation}
2=4\pi\int_0^\infty x^2n_f\,\mathrm{d}x.
\end{equation}

\subsection{i. Small-$\boldsymbol{m}$ asymptotics}

For $m\ll 1$, we can begin by performing a small-$x$ expansion on the above expressions. Strictly speaking, it is not guaranteed that $m\ll1$ will imply $x\ll1$, since the typical radial extent of the solution could increase to compensate as $m$ decreases. From our numerical results, however, we observe that in fact the opposite occurs, with the solution becoming compressed towards $r=0$ as $m$ decreases; hence we are justified in equating small $m$ with small $x$. It is then straightforward to perform the asymptotic analysis on $n_f$ and $T$, obtaining:
\begin{align}
     x^2n_f&\approx \frac{1}{2\pi Gm^2}\left[1+\frac{2x^2}{h_m}\right]^{-1};\\
    T&\approx \left[1+\frac{h_m}{2x^2}\right]^{-1}.
\end{align}
Note that we cannot remove any further terms from these expressions, since we do not yet know the $m$-dependence of $c_m$. Using the asymptotic form for the fermion density, it is then possible to solve the normalization condition analytically, from which we can obtain the leading-order value of $c_m$:
\begin{align}
    1&=\left(\frac{4\pi c_m^2}{m^3G}\right)^{1/2}\left[\tan^{-1}\left(\sqrt{2}xh_m^{-1/2}\right)\right]^\infty_0\\
    \implies c_m&=\left(\frac{m^3G}{\pi^3}\right)^{1/2}.\label{eqSuppCMsmallM}
\end{align}
Substituting this back into our asymptotic forms for $n_f$ and $T$, we obtain:
\begin{align}
    x^2n_f&\approx \frac{1}{2\pi Gm^2}\left[1+\left(\frac{\pi x}{2Gm^2}\right)^2\right]^{-1}; \\
    T&\approx \left[1+\left(\frac{2Gm^2}{\pi x}\right)^2\right]^{-1}.
\end{align}
These expressions reflect those stated in the main paper, and represent the `solid ball' profile exhibited by the fermion density at small $m$.

\begin{figure}
	\includegraphics{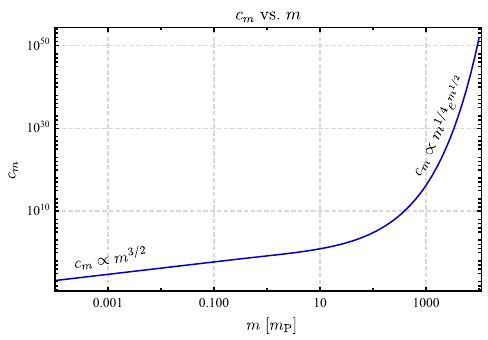}
	\caption{The numerically determined value of the dimensionless parameter $c_m$ as a function of fermion mass $m$, showing the transition from power-law behavior to exponential increase, in line with the analytic asymptotic analysis.}
	\label{figCm}
\end{figure}

Furthermore, it is straightforward to infer a characteristic radius for the solution, $x=2Gm^2/\pi$, agreeing with the small-$m$ behavior for the radial extent $R$ shown in Fig.~\ref{figMassRadius} (recalling that $x=mr$). For completeness, we note that the asymptotic value for $c_m$ (\ref{eqSuppCMsmallM}) is also in agreement with the numerical results, as can be seen from Fig.~\ref{figCm}.

\vspace{-5pt}
\subsection{ii. Large-$\boldsymbol{m}$ asymptotics}
\vspace{-5pt}

Performing an asymptotic analysis for large $m$ proves somewhat less straightforward. In what follows, we shall therefore make use of knowledge informed from our numerical solutions, in order to make progress analytically. First, we note that the radial extent of the solutions at large $m$ behaves like $R\sim m^{-1/2}$, and hence we can safely equate large values of $m$ with large values of $x$ and perform a large-$x$ expansion on (\ref{eqSuppNfMaster}) and (\ref{eqSuppTMaster}), obtaining:
\begin{align}
    x^2n_f&\approx \frac{8c_m^2m^{-1}x^3}{2x^3e^{2x}+h_m};\\
    T&\approx \left[1+\frac{h_m}{2x^3e^{2x}}\right]^{-1}.
\end{align}
As noted earlier, this cannot be simplified further without knowing the $m$-dependence of $h_m$. To obtain this, we are required to solve the normalization condition, which approximates to the following:
\begin{align}
    1=\frac{16\pi c_m^2}{m} \int_0^\infty \frac{x^3}{2x^3e^{2x}+h_m} \,\mathrm{d}x .\label{eqSuppLargeMInt}
\end{align}
This integral does not appear to have an analytic solution, and we must therefore make a further approximation to simplify the problem. Although techniques such as the method of steepest descents appear promising, these ultimately fail since the fermion density is highly non-Gaussian at large $m$. Instead, we take the approach of splitting the integral, noting that the integrand contains a peak (denoted $x_p$) either side of which each respective term in the denominator is dominant, allowing us to write:
\begin{align}
    1&\approx\frac{16\pi c_m^2}{m}\left(\int_0^{x_p} \frac{x^3}{h_m}\,\mathrm{d}x+\int_{x_p}^\infty \frac{1}{2e^{2x}}\,\mathrm{d}x\right)\\
    &\approx \frac{4\pi c_m^2}{m}
    \left(h_m^{-1}x_p^4+e^{-2x_p}\right).\label{eqSupp67}
\end{align}

The radius of the peak, $x_p$, can be obtained by simply differentiating the integrand of (\ref{eqSuppLargeMInt}) to obtain:
\begin{align}
 x_p=2W_0\left[\left(\frac{3h_m}{64}\right)^{1/4}\right],
\end{align}
where $W_0$ is the Lambert $W$ function. Noting that $W_0(z) \approx \log(z)$ for large $z$, we can therefore neglect the second term in (\ref{eqSupp67}) at large $m$, since it will be of order $1/m$. This is in line with the numerical observation that the fermion density becomes increasingly non-Gaussian as $m$ increases. It is then straightforward to solve the normalization condition for $c_m$ to obtain:
\begin{align}
    c_m=\sqrt{\frac{m}{3\pi}}\exp\left[ \left( 2Gm^2 \right)^{1/4}\right].
\end{align}
While we find that the exponential dependence exhibited by this expression does indeed agree with our numerical results, the power of $m$ in the pre-factor does not. This is most likely a consequence of the approximation involved when splitting the integral. The correct $m$-dependence can, however, be inferred using the following argument. First, assume that the leading-order $m$ dependence of $c_m$ is $c_m \sim m^{a}\exp(m^{1/2})$, where $a$ is to be determined. We can surmise that, at large $m$, the two terms in the denominator of (\ref{eqSuppLargeMInt}) must be of the same order, since otherwise the solution would not exhibit the (numerically) observed radial profile. Defining a typical radius within the solution as $x_0$, we can therefore write:
\begin{align}
    2x_0^3e^{2x_0}&\sim h_m \sim G m^{2a+1}e^{2m^{1/2}}.
\end{align}
Equating terms implies $x_0\sim m^{1/2}$ and $a=1/4$, and thus
\begin{align}
    c_m\sim \left(m^2/G\right)^{1/8}\exp\left[ \left( 2Gm^2 \right)^{1/4}\right],
\end{align}
where we have used dimensional analysis to infer the dependence on $G$. This indeed agrees with the numerical relationship shown in Fig.~\ref{figCm}.

Finally, we can substitute this form back into our expression for the peak radius $x_p$ to obtain:
\begin{align}
x_p&\sim 2 W_0\left[\left(\frac{3\pi Gc_m^2m}{8}\right)^{1/4}\right]\\
&\sim 2\log \left(G^{1/4}m^{5/4}\right)+\left(\frac{Gm^2}{8}\right)^{1/4}\\
&\sim \left(Gm^2\right)^{1/4}.
\end{align}
Using this radius as an approximate measure for the radial extent of the solution $R$, we conclude that $R\sim m^{-1/2}$ (recalling that $x=mr$), which is precisely the numerically-obtained relationship shown previously in Fig.~\ref{figMassRadius}.

\section{Energy conditions}

We stated at the end of the main paper that the stealth solution violates a number of the usual energy conditions that are known to apply to non-exotic classical matter. For completeness, we here present the full expressions for these. The relevant quantities are the energy density $\rho$, radial pressure $P_r$, and azimuthal pressure $P_\perp$, all of which are defined via the stress-energy tensor $T^\mu_{\;\;\nu}$:
\begin{align}
    &\rho=-T^t_{\;\;t}\,;
    &&P_r=T^r_{\;\;r}\,;
    &&P_\perp=T^\theta_{\;\;\theta}=T^\phi_{\;\;\phi}.
\end{align}
For the Einstein--Dirac system, the stress-energy tensor takes the following general form:
\begin{align}
    T^{\mu}_{\;\;\nu}=-\sum_{a}\Re\left\{\overline{\Psi}_a\left(i\gamma^\mu\partial_\nu\right)\Psi_a\right\}
\end{align}
from which we obtain:
\begin{align}
    \rho&=0;
    \\P_r&=\frac{-2c_m^2m^3e^{-2x}}{2x^4+h_mx^2\left[(1-2x)e^{-2x}-4x^2\mathrm{Ei}(-2x)\right]};
    \\P_\perp&=\frac{4c_m^2m^3(1+x)e^{-2x}}{2x^4+h_mx^2\left[(1-2x)e^{-2x}-4x^2\mathrm{Ei}(-2x)\right]}.
\end{align}
Note that the radial pressure is negative for all radii (since $(1-2x)e^{-2x}-4x^2\mathrm{Ei}(-2x)$ is strictly positive for $x>0$), consistent with the stealth solution being gravitationally repulsive. 

As a result of this negative radial pressure, combined with the zero energy density (a consequence of $\omega=0$), the stealth solution violates both the weak and strong energy conditions, since $\rho+P_r<0$. The dominant energy condition is also violated, since $\rho < |P_r|$ and $\rho < |P_\perp|$. This is usually interpreted as allowing the speed of energy flow of matter to be superluminal. As mentioned in the main paper, however, it is debatable whether these conditions should apply to quantum mechanical systems such as the one considered here.

\end{document}